\documentclass[runningheads]{llncs}
\usepackage{booktabs} 
\usepackage{graphicx}
\usepackage{multirow}
\usepackage{amsfonts}        
\usepackage{nicefrac}        
\usepackage{microtype}       
\usepackage{subfigure}
\usepackage{enumitem}

\usepackage{algorithm}
\usepackage{algorithmic}

\usepackage{amsmath}
\usepackage{amssymb}
\usepackage{color}
\usepackage{float}
\usepackage{bbding}
\newcommand{\tabincell}[2]{\begin{tabular}{@{}#1@{}}#2\end{tabular}}
\begin{document}
\title{Automatic Data Augmentation for 3D Medical Image Segmentation}
%
%
\author{Ju Xu\inst{1 \thanks{Equal contributions.} } \and
Mengzhang Li\inst{1,2 \footnotemark[1]} \and
Zhanxing Zhu\inst{3,1}}

\institute{Center for Data Science, Peking University, Beijing, China \and Canon Medical Systems, Beijing, China \and School of Mathematical Sciences, Peking University, Beijing, China\\
\email{\{xuju, mcmong, zhanxing.zhu\}@pku.edu.cn}}
\maketitle              
\begin{abstract}
Data augmentation is an effective and universal technique for improving generalization performance of deep neural networks. It could enrich diversity of training samples that is essential in medical image segmentation tasks because 1) the scale of medical image dataset is typically smaller, which may increase the risk of overfitting; 2) the shape and modality of different objects such as organs or tumors are unique, thus requiring customized data augmentation policy. However, most data augmentation implementations are hand-crafted and suboptimal in medical image processing. To fully exploit the potential of data augmentation, we propose an efficient algorithm to automatically search for the optimal augmentation strategies. We formulate the coupled optimization w.r.t. network weights and augmentation parameters into a differentiable form by means of stochastic relaxation. This formulation allows us to apply alternative gradient-based methods to solve it, i.e. stochastic natural gradient method with adaptive step-size. To the best of our knowledge, it is the first time that differentiable automatic data augmentation is employed in medical image segmentation tasks. Our numerical experiments demonstrate that the proposed approach significantly  outperforms existing build-in data augmentation of state-of-the-art models.

\keywords{Medical Image Segmentation  \and Data Augmentation \and AutoML}
\end{abstract}
\section{Introduction}
In the past few years, deep neural network has achieved incredible progress in medical image segmentation tasks   and promoted booming development of computer assisted intervention.  This has benefitted  research and clinical treatment of disease diagnosis, treatment design and prognosis evaluation~\cite{ronneberger2015u,tajbakhsh2016convolutional}. Given the training data, researchers proposed various 2D/3D medical image segmentation models for supervised or semi-supervised tasks~\cite{isensee2019nnu,ganaye2019removing}. However, the performance of deep learning models heavily relies on large  scale well-labeled data. Currently, data augmentation is a widely used and effective technique to increase the amount and diversity of available data, and thus improving models' generalization performance. In the domain of natural image processing, typical data augmentation strategies include \emph{manually} cropping, rotating or adding random noise to the original images. Besides thess ad-hoc approaches,  generative models~\cite{huang2018auggan} and unsupervised learning models~\cite{xie2019unsupervised} are also employed for generating extra data. Unfortunately, those augmentation techniques might not be optimal for a specific task, and thus the customized data augmentation strategy is required. Recently, researchers proposed to  search the augmentation policy by reinforcement learning~\cite{cubuk2019autoaugment} or density matching~\cite{lim2019fast}, inspired by previously works of automatic machine learning (AutoML) on neural architecture search (NAS~\cite{liu2018darts,pham2018efficient,zoph2018learning}). 

For medical image segmentation tasks,  data augmentation techniques are also used in UNet and  its variants nnUNet~\cite{isensee2019nnu}, R2U-Net~\cite{alom2018recurrent}, etc. However, these methods are simple and hand-made, and the  improvement of segmentation accuracy is limited. In \cite{yang2019searching}, the authors proposed to utilize reinforcement learning to search for augmentation strategies. However, it costs 768 GPU hours and it only searchs the probability of each augmentation strategy in \cite{yang2019searching}. 
Moreover, the difference between natural and medical images such as spatial contextual correlation,  smaller scale of dataset and unique pattern of specified organs or tumor makes the augmentation strategies adopted in natural images difficult be transferred to medical domains. 

In this paper, we propose an \emph{automatic data augmentation} framework (ASNG) through searching the optimal augmentation policy, particularly for 3D medical image segmentation tasks. It’s the first automatic data augmentation work in whole semantic segmentation filed. The contributions of our paper are as follows: 
\begin{itemize}
\item It’s the first time that we formulate the auto-augmentation problem into a bi-level optimization problem and apply an approximate algorithm to solve it
\item The designed search space in medical image field is novel. Different from previous methods which searched for a fixed magnitude of operations, we search for an interval of magnitude
\item Different from previous method which searched for a fixed augmentation strategy, the searched augmentation strategy of our method is dynamically changing during the training. Besides, we don’t need to retrain the target network after the searching process
\item Experiments demonstrate that our ASNG can indeed achieve the SOTA of the performance
\end{itemize}

\section{Method}

\begin{figure}[t!]
\centering
\includegraphics[width=0.8\linewidth]{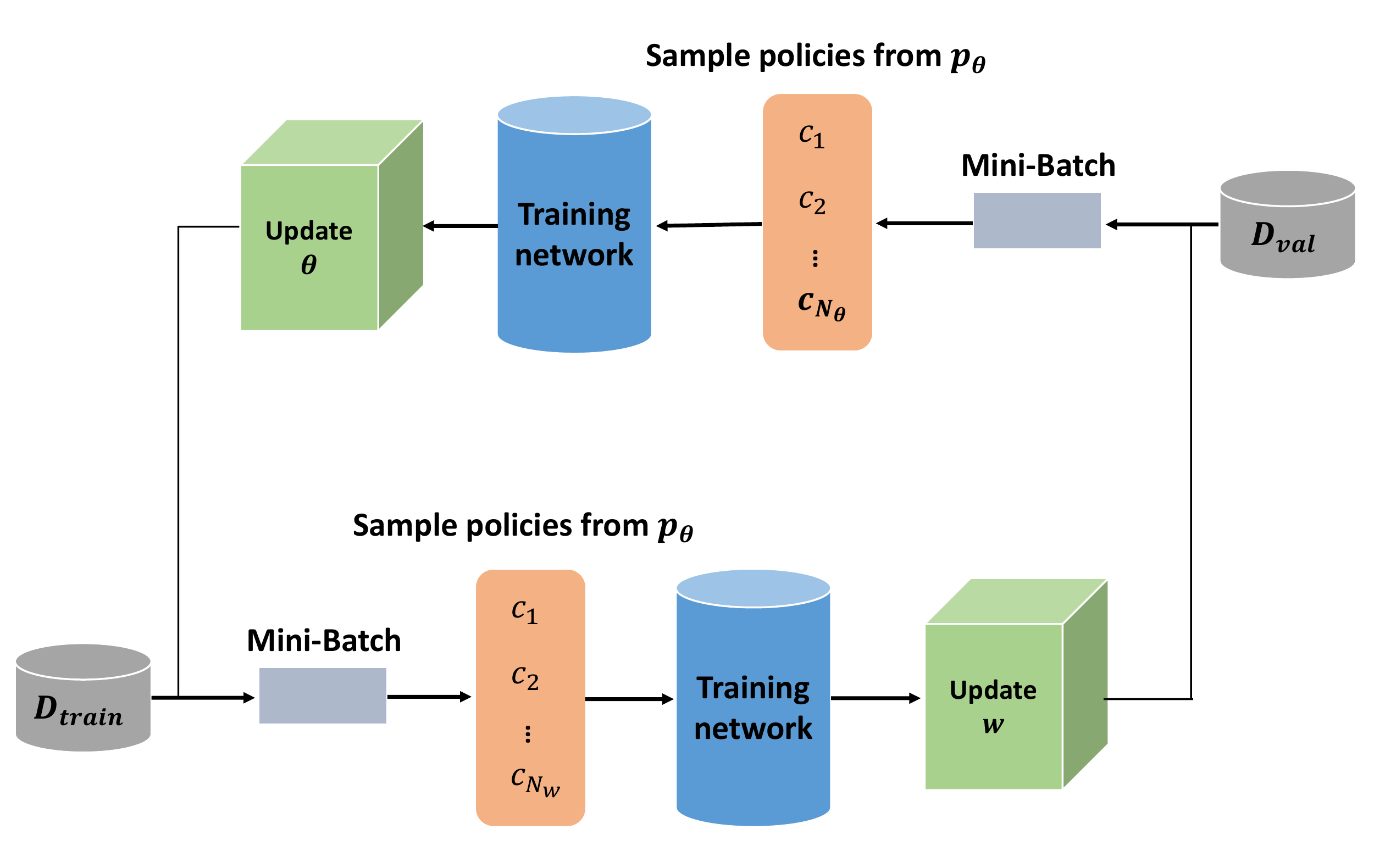}
\caption{The framework of our proposed method. $D_{train}$, $D_{val}$ represent training dataset, validation dataset, respectively. $p_{\theta}$ is the distribution of $c$.}
\label{framework}
\end{figure}

In our method, we formulate the problem of finding the optimal augmentation policy as a discrete search problem. Our method consists of two components: the designed of search space and search algorithm. The search algorithm samples a data augmentation policy $S$ from the search space consisting of proposed operations, and then decides the magnitude of the operation and the probability of applying this operation. The framework of our method can be seen in Fig.\ref{framework}. We will elaborate the two components in the following.

\subsection{Search Space of Data Augmentation for 3D Medical Images }
Since it is the first work for applying AutoAugment strategies in medical image area, we have to design the search space for our ASNG algorithm. In our search space, a policy consists of seven image operations to be applied in a sequential manner. Each image operation is associated with two hyperparameters: 1) the probability of applying this operation, and 2) the interval of magnitude for the image operation.

\begin{figure}
	\centering
	\includegraphics[width=0.8\linewidth]{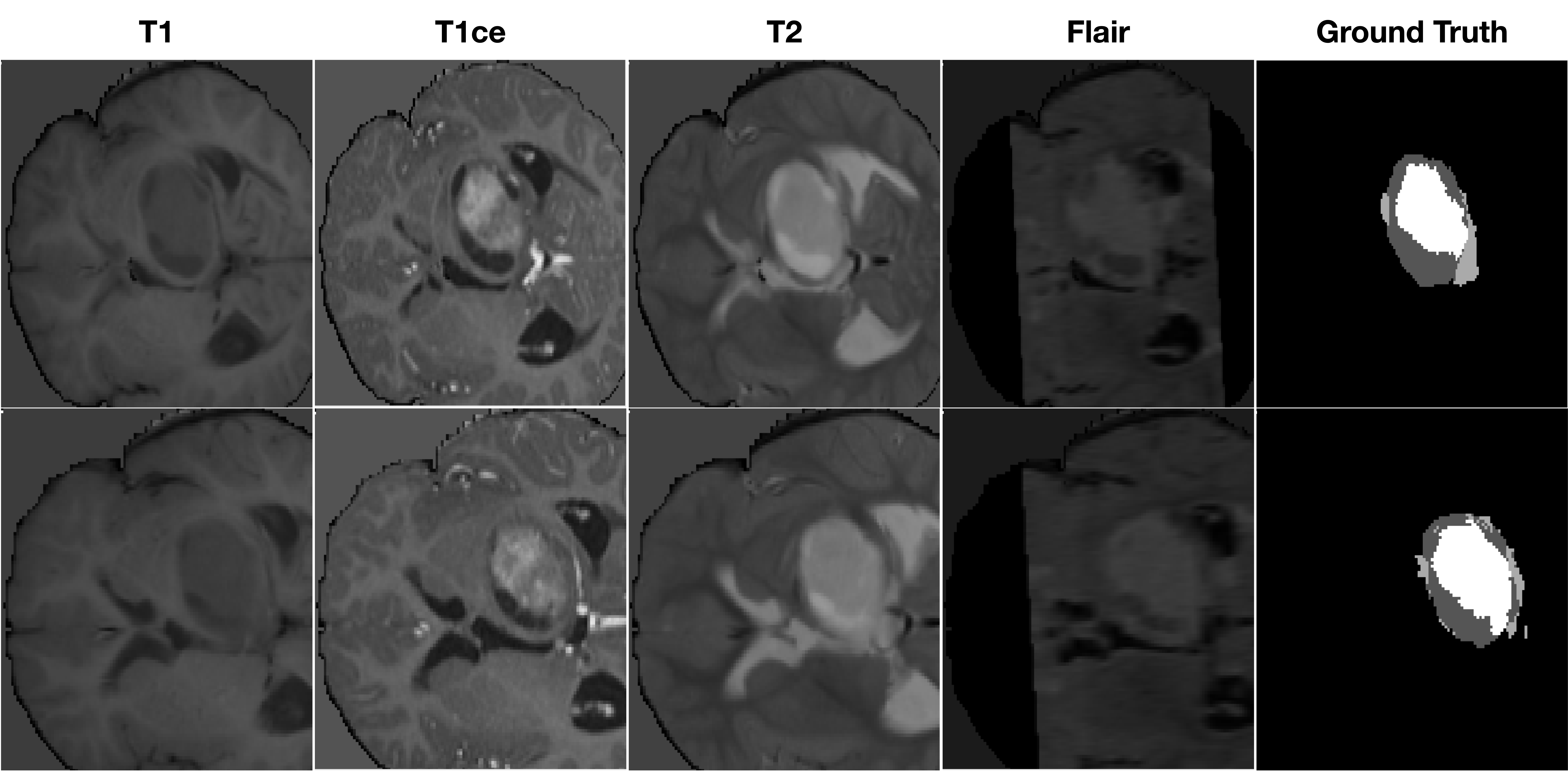}
	\caption{Visualization of the proposed data augmentation on certain 2D section of Task01 BrainTumour dataset. \textbf{Top:} original image and label, \textbf{Bottom:} the image and label generated by the searched data augmentation policy, in which the operations are from  Table \ref{tab:para}.}
	\label{fig:sample_aug}
\end{figure}

The seven image operations we used in our experiments are from batchgenerators, a pupular Python image library\footnote{https://github.com/MIC-DKFZ/batchgenerators}, including Scale, RoateX/Y/Z, Alpha (magnitude of the elastic deformation), Sigma (scale of the elastic deformation), Gamma (same as gamma correction in photos or computer monitors). In order to increase the diversity of augmentation policies, we do not fix a specific magnitude for an operation like previous works~\cite{cubuk2019autoaugment}, but set an interval of magnitude for an operation. Therefore we should decide the left boundary of interval (LB) and the right one (RB). To decrease the search complexity, we discretize the range of magnitude into 11 values with uniform spacing so that we can use a discrete search algorithm to find them. Besides the magnitude of transformation operation, we also search for the probability of conducting these transformations, i.e. the probability of applying scale transformation, rotation, gamma transformation, and elastic deformation, denoted as $p_{scale}$, $p_{rot}$,  $p_{gamma}$,  $p_{eldef}$, respectively. Similarly, we also discretize the probability of applying that operation into 11 values with uniform spacing. Table \ref{tab:para} summarizes the range of magnitudes and possibilities for the seven operations. Fig.~\ref{fig:sample_aug} shows one example of augmented  image and label based on our method, in which the image transformations are from the defined search space.  

We can easily observe that naively searching one augmentation strategy becomes a search problem with $11^{11}$ possibilities. The search space is so huge that an efficient algorithm is required, as proposed in the following.

\begin{table}
\centering
\caption{The range of parameters in strategies we will search.}
\begin{tabular}{p{2cm}p{2cm}p{2cm}p{2cm}c}
\toprule
Operation & LB & RB & Probability & Range\\
\toprule
Scale & [0.5, 1.0] & [1.0, 1.5] & $p_{scale}$ & [0, 1] \\
RotationX & [$\frac{-\pi}{6}$, 0] & [0, $\frac{-\pi}{6}$] & $p_{rot}$ & [0, 1]\\
RotationY & [$\frac{-\pi}{6}$, 0] & [0, $\frac{-\pi}{6}$] & $p_{rot}$ & [0, 1]\\
RotationZ & [$\frac{-\pi}{6}$, 0] & [0, $\frac{-\pi}{6}$] & $p_{rot}$ & [0, 1]\\
Alpha & [0, 450] & [450, 900] & $p_{eldef}$ & [0, 1] \\
Sigma & [0, 7] & [7, 14] & $p_{eldef}$ & [0, 1] \\
Gamma & [0.5, 1] & [1, 1.5]  & $p_{gamma}$ & [0, 1] \\
\bottomrule
\end{tabular}
\label{tab:para}
\end{table}

\subsection{Stochastic Relaxation Optimization of Policy Sampling}

We denote $f(w, c)$ as the objective function, where $w \in \mathcal{W}$ are network parameters and $c \in \mathcal{C}$ are data augmentation strategies.
$f_{train}$ and $f_{val}$ are the training and the validation loss, respectively. Both losses are determined not only by the augmentation policy $c$, but also the weight $w$. The goal for augmentation strategy search is to find $c^*$ that minimizes the validation loss $f_{val}(w^*, c^*)$, where the weights are obtained  by minimizing the training loss $w^* = \text{argmin}_w f_{train}(w, c^*)$. Thus augmentation strategy search is a bi-level optimization problem, we can write the problem as follows:
\begin{align}
    &\min_c f_{val}(w^*(c), c) \\
    &s.t. \quad w^*(c) = \mathop{\text{argmin}}\limits_w f_{train}(w,c)
    \label{opt:2}
\end{align}

Solving the above problem is not easy, since we cannot obtain the gradient w.r.t. $c$, thus it is hard to optimize $c$ via gradient descent. Though simple grid search or reinforcement learning proposed in~\cite{cubuk2019autoaugment} can be utilized to search for $c$, the computational cost is extremely high if we evaluate the performance of every $c$. To this end, we propose to solve this optimization problem efficiently first by stochastic relaxation and then applying natural gradient descent~\cite{amari1998natural}, as described in the following. 

\subsubsection{Stochastic Relaxation}
We turn the original optimization problem into an optimization of differentiable objective $J$ by stochastic relaxation~\cite{akimoto2019adaptive}. The basic idea of stochastic relaxation is: instead of directly optimizing w.r.t $c$, we consider a distribution $p_{\theta}(c)$ over $c$ parametrized by $\theta$, and minimize the expected value of the validation loss $f_{val}$ w.r.t $\theta$, i.e.,
\begin{align}
    &\min\limits_{\theta} J(\theta) = \int_{c\in \mathcal{C}} f_{val}(w^*(c), c)p_{\theta}(c) dc\\
     &s.t. \quad w^*(c) = \mathop{\text{argmin}}\limits_w f_{train}(w,c)
\end{align}

%
The stochastic relaxation makes $J$ differentiable w.r.t both $w$ and $\theta$. Therefore we can update $w$ and $\theta$ by gradient descent. However, the gradient $\nabla_w J(w, \theta)$ is not tractable because we can not evaluate the mean performance of $c \in \mathcal{C}$ in a closed-form way. Here we estimate the gradient by Monte-Carlo (MC) using $\nabla_{w} J(w^t, c_i)$ with i.i.d. samples $c_i \sim p_{\theta^t}(c),\, i=1,\dots, N_w$, namely:
\begin{align}
    G_w(w^t, \theta^t) = \frac{1}{N_w} \sum\limits_{i=1}^{N_w} \nabla_w f_{train}(w^t, c_i)
\end{align}

Now we can approximate  $\nabla_w J(w, \theta)$ with the stochastic gradient $G_w(w^t, \theta^t)$, $w^t$ can be updated as
\begin{align}
    w^{t+1} = w^t - \epsilon_w G_w(w^t, \theta^t),
    \label{update:w}
\end{align}
where $\epsilon_w$  is the learning rate for network parameters.
Due to that the distance between two probability distribution is not Euclidean, updating $\theta$ directly by gradient descent like $w$ is not appropriate. We then resort to natural gradient (NG \cite{amari1998natural}) designed for parametric probability distributions, 
\begin{align}
    \theta^{t+1} = \theta^t - \epsilon_{\theta} F(\theta_t)^{-1} \nabla_{\theta} J(w, \theta),
\end{align}
where $F(\theta_t)$ is the Fisher matrix,  $\epsilon_{\theta}$ is the learning rate. The probability distribution we utilized for $c \in \mathcal{C}$ is multinomial distribution. How to calculate the Fisher matrix can be seen in \cite{akimoto2019adaptive}.
 Similiar with \cite{akimoto2019adaptive}, we utilize adaptive step-size $\epsilon_{\theta}$ to make the learning process faster. Monte-Carlo is also adopted to approximate $\nabla_{\theta} J(w, \theta)$, and then
\begin{align}
\theta^{t+1} = \theta^t - \epsilon_{\theta} F(\theta_t)^{-1} \frac{1}{N_{\theta}} \sum\limits_{j=1}^{N_{\theta}} \nabla_{\theta}f_{val}(w^{t+1}, c_j) \ln p_{\theta}(c_j)
\label{update:theta}
\end{align}
We summarize the procedure of our proposed approach in Algorithm~\ref{alg:asng}.

\begin{algorithm}[ht]
	\caption{ASNG}
	\small
	\begin{algorithmic}[1]
        \STATE \textbf{Input: } $w^0$, $\theta^0$, $\epsilon_w$. $\epsilon_{\theta}$, $N_w$, $N_{\theta}$
        \STATE \textbf{Input: } Training dataset $D_{train}$, validation dataset $D_{val}$, test dataset $D_{test}$.
        \FOR{i=1 to epoch}
        \FOR{t=1 to T}
            \STATE Generate $N_w$ policys according to $p_{\theta_t}$
            \STATE Augment training data from $D_{train}$ with $N_w$ policys, respectively;
            \STATE Obtain the loss $f_{train}(w_t, c_i)$ ($i=1,\dots, N_w$) on $D_{train}$;
            \STATE Update $w_t$ according to Equation~\ref{update:w}, then obtain $w_{t+1}$;
            \STATE Generate $N_{\theta}$ policys according to $p_{\theta_t}$; \\
            \FOR{j=1 to $N_{\theta}$}
                \STATE Augment training data according to policy $c_j$;
                \STATE Update $w_t$ to obtain $\hat{w_t}$;
                \STATE Obtain the validation loss $f_{val}(\hat{w_t})^j$  on $D_{val}$;
                \STATE Restore the network parameters, $\hat{w_t} = w_t$;
            \ENDFOR
            \STATE Utilize validation loss $f_{val}(\hat{w_t})^j$, policys $c_j$  ($j=1,\dots, N_{\theta}$) to update $\theta_t$ according to equation~\ref{update:theta};
            
        \ENDFOR
        \ENDFOR
       \STATE  Test the network on $D_{test}$;
		\RETURN final networks.
	\end{algorithmic}
	\label{alg:asng}
\end{algorithm}
\section{Implementation  and Experiments}
\subsection{Datasets and Implementation Details}
\textbf{Datasets}: We conduct the proposed  method on three 3D segmentation tasks used in the medical segmentation decathlon challenge (MSD\footnote{http://medicaldecathlon.com/}): (1) Task01 Brain Tumour (484 labeled images, 3 classes), (2) Task02 Heart (20 labeled images, 1 class) and (3) Task05 Prostate (32 labeled images, 2 classes). Each dataset is collected for a specified task, their various input sizes, voxel spacings and foreground targets are suitable for demonstrating our algorithm's generalization. We evaluate the performance with 5-fold cross validation as \cite{bae2019resource,kim2019scalable} since the ground truth labels for test dataset are not publicly available. \\
\textbf{Compared Methods}  include 3D U-ResNet~\cite{yu2017volumetric}, SCNAS~\cite{kim2019scalable}, nnUNet without data augmentation (nnUNet NoDA) and 3D nnUNet~\cite{isensee2019nnu} with default data augmentation strategy\footnote{https://github.com/MIC-DKFZ/nnUNet/}. 3D U-ResNet utilizes residual blocks and attention gates; and SCNAS is the latest neural architecture search model for 3D medical image segmentation, which applies a scalable gradient-based optimization to find the optimal model architecture. The method proposed in SCNAS can't utilized for differentiable autoaugmentation strategies search. In SCNAS, a mixed operation is created by adding all these operations in search space based on the importance of each operation. However, we can’t add the transformation results of each augmentation strategy. There we don't apply the proposed method of SCNAS to our augmentation strategies search tasks.  Note that the code of nnUnet has already implemented random augmentation. In nnUnet, the magnitude of operation is sampled from a predefined interval in every training epoch. AutoAugment~\cite{cubuk2019autoaugment} costs 5000 GPU hours to search for a policy. FastAutoAugment~\cite{lim2019fast} needs to spilt the training dataset as K folds. However, dataset in medical image area is quite small. Training model in a small dataset will overfit. Therefore, we don't compare our method with AutoAugment and FastAutoAugment. The prediction result is inferenced using a sliding window with half the patch size ensuring 50\% overlapping, i.e., each voxel is inferenced at least two times at test.  \\ 
\textbf{Implementation Details} We preprocess the data with same pipeline used in 3D nnUNet. We unify the identical voxel spacing values by proper interpolation due to different spacing values of each case, i.e. resampling them to 0.7 mm $\times$ 0.7 mm $\times$ 0.7 mm firstly. We apply Z-score normalization of voxel value for each input channel separately; and grip the input patch whose size is set as 128 $\times$ 128 $\times$ 128, and its foreground ratio is set larger than 1/3 ensuring UNet variants could learn features of foreground. 

Following the default data augmentation policy in nnUNet, we utilize scaling, rotation, elastic and gamma transformation both in training and test. The parameters of the operation and probability of conducting that operation are both within search space of our ASNG algorithm.  

The code is implemented using PyTorch 1.0.0. The ADAM optimizer is utilized for training where the learning rate and  weight decay  are initialized as $3\times 10^{-4}$ and $3\times 10^{-5}$, respectively, where it is reduced by 80\% if the training loss is not reduced over 30 epochs. Besides ASNG, the training process of other benchmarks would last for 500 epochs if the learning rate is larger than $10^{-7}$. Following \cite{kim2019scalable} and \cite{isensee2019nnu}, the loss function for 3D U-ResNet and SCNAS is Jaccard distance, for nnUNet and ASNG is sum of minus Dice similarity and Cross Entropy. Considering the training time, ASNG is trained for 50, 200 and 200 epochs on Brain Tumour, Heart and Prostate, respectively, with batch size of 2. It takes about 10 days on one NVIDIA TITAN RTX GPU, compared with that one integrated nnUNet training procedure takes about 3 days. The sampling times $T$ of ASNG is 2 because of the limited memory of GPU, though larger $T$ could produce better numerical results. Our codes can be found here \footnote{https://github.com/MengzhangLI/ASNG}.

\subsection{Experimental Results}
Our result is shown in Table \ref{tab1}. Because of unavailable labels of test set and restricted online submission times, those Auto ML models on 3D Medical Image Segmentation tasks are all evaluated on validation set. In this paper we still follows this metric for fair comparison. ASNG outperforms other architectures especially 3D nnUNet, which is the best medical image segmentation framework with default data augmentation. It should be noted that since Heart and Prostate only have 20 and 32 labeled images, in~\cite{kim2019scalable} the  obtained architecture of SCNAS based  on the first fold of 484 labeled Brain Tumour images is transferred to Heart and Prostate tasks to avoid overfitting. Remarkably, our method, applied only on the basic network architecture, could still achieve best prediction accuracy. This clearly demonstrates the necessity and effectiveness of data augmentation policy search in 3D medical image segmentation.

\begin{figure}[t!]
	\centering
	\includegraphics[width=1\linewidth]{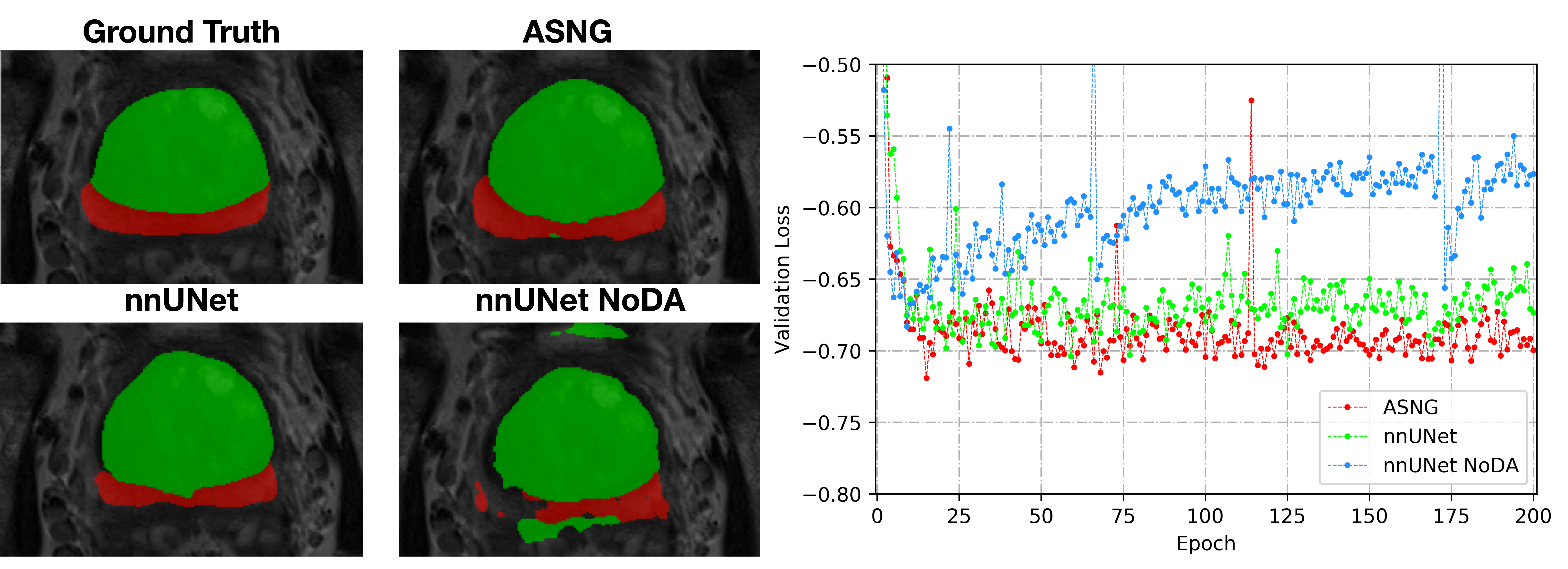}
	\caption{Results on Task05 Prostate of selected architectures. \textbf{Left:} Example of inference, green mask represents peripheral zone and red mask represents transitional zone. \textbf{Right:} The trend of loss on validation set.}
	\label{fig2}
\end{figure}

Figure \ref{fig2} shows the example of segmentations results and validation loss w.r.t. number of epochs in the Prostate task. We can observe that our method ASNG can produce better prediction and  achieve more stable improvement during training than other compared methods. 

In this paper~\cite{yang2019searching}, the proposed method utilizes reinforcement learning to search for augmentation strategies, which costs 768 GPU hours while our method costs less than 100. And their result (Dice 0.92) on task 02 is worse than ours (0.933). Besides, the first paper only searched the probability of each augmentation strategy while our method not only search the probability but also the magnitude.

\begin{table}[t!]
	\scalebox{0.85}{
	\begin{tabular}{|c|c|c|c|c|c|c|c|c|}
		\hline
		\multirow{2}{*}{Label} & \multicolumn{4}{c|}{Brain Tumour}                                 & Heart          & \multicolumn{3}{c|}{Prostate}                    \\ \cline{2-9} 
		& Edema          & \tabincell{c}{Non-\\Enhancing}  & Enhancing      & Average        & Left atrium    & Peripheral     & Transitional   & Average        \\ \hline
		U-ResNet            & 79.10          & 58.38          & 77.37          & 71.61          & 91.48          & 48.37          & 79.17          & 63.77          \\ \hline
		nnUNet NoDA         & 81.27          & 60.92          & 77.90          & 73.36          & 92.85          & 58.61          & 83.61          & 71.11          \\ \hline
		nnUNet              & 81.68          & 61.29          & 77.97          & 73.65          & 93.21          & 63.14          & 86.53          & 74.84          \\ \hline
		SCNAS                  & 80.41          & 59.85          & 78.50          & 72.92          & 91.91          & 53.81          & 82.02          & 67.92          \\ \hline
		ASNG                   & \textbf{81.94} & \textbf{61.85} & \textbf{79.35} & \textbf{74.38} & \textbf{93.27} & \textbf{67.40} & \textbf{87.05} & \textbf{77.22} \\ \hline
	\end{tabular}}
	\caption{Average Dice similarity coefficients (\%) for Brain tumor, Heart, and Prostate 3D segmentation tasks of MSD. }
	\label{tab1}
	\vspace{-5mm}
\end{table}
\section{Conclusion}
We have proposed an automatic data augmentation strategy to accommodate 3D medical image segmentation tasks. By configuring proper search space followed by  gradient-based optimization, the  customized data augmentation strategy for each task could be obtained. The numerical results for different segmentation tasks show it could outperform the state-of-the-art models that are widely used in this area.  Furthermore, the proposed approach shows that, compared with searching network architectures, searching for optimal data augmentation policy is also important. As for future work, designing better search space and accelerating the search process can be considered.

%
%
%
%
%
 \bibliographystyle{splncs04}
 \bibliography{paper921}
%
%
%
%
%
\end{document}